\documentstyle[12pt,epsfig]{article} 
\begin{document}


\def\beq             {\begin{equation}}
\def\eeq             {\end{equation}}
\def\beqd            {\begin{displaymath}}
\def\eeqd            {\end{displaymath}}
\def\baa             {\begin{array}}
\def\eaa             {\end{array}}
\def\beqaa           {\begin{eqnarray}}
\def\eeqaa           {\end{eqnarray}}
\def\beqaad          {\begin{eqnarray*}}
\def\eeqaad          {\end{eqnarray*}}
\def\btabu           {\begin{tabular}}
\def\etabu           {\end{tabular}}
\def\bfig            {\begin{figure}}
\def\efig            {\end{figure}}
\def\bce             {\begin{center}}
\def\ece             {\end{center}}

\def\ind             {\indent}
\def\noi             {\noindent}
\def\nn              {\nonumber}

\newcommand{\eq}[1]  {\mbox{eq.~(\ref{eq:#1})}}
\newcommand{\fig}[1] {\mbox{Fig.~\ref{fig:#1}}}   

\def\lbr             {\lbrack}
\def\rbr             {\rbrack}
\def\ti              {\tilde}
\def\q               {\bar}

\def\a               {\alpha}
\def\b               {\beta}
\def\g               {\gamma}
\def\G               {\Gamma}
\def\l               {\lambda}
\def\t               {\theta}
\def\s               {\sigma}
\def\x               {\chi}

\def\ee              {$e^+ e^-$}
\def\eeto            {e^+ e^- \to}

\def\sq              {\ti q}
\def\sqsq            {\sq_1\,\bar{\sq}_1}
\def\slep            {\ti\ell}

\def\st              {\ti t}
\def\stij            {\ti t_{i,i}}
\def\stst            {\st_1\,\bar{\st}_1}
\def\stistj          {\st_i\,\bar{\st}_j}

\def\sb              {\ti b}
\def\sbij            {\sb_{i,j}}
\def\sbsb            {\sb_1\,\bar{\sb}_1}
\def\sbisbj          {\sb_i\,\bar{\sb}_j}

\def\stau            {\ti \tau}
\def\stauij          {\stau_{i,i}}
\def\staustau        {\stau_1\,\bar{\stau}_1}
\def\stauistauj      {\stau_i\,\bar{\stau}_j}

\def\sf              {\ti f}
\def\sfij            {\ti f_{i,i}}
\def\sfsf            {\sf_1\,\bar{\sf}_1}
\def\sfisfj          {\sf_i\,\bar{\sf}_j}

\def\ch              {\ti \x^\pm}
\def\chp             {\ti \x^+}
\def\chm             {\ti \x^-}
\def\nt              {\ti \x^0}

\newcommand{\mst}[1]   {m_{\st_{#1}}}
\newcommand{\msb}[1]   {m_{\sb_{#1}}}
\newcommand{\mstau}[1] {m_{\stau_{#1}}}
\newcommand{\msf}[1]   {m_{\sf_{#1}}}
\newcommand{\mch}[1]   {m_{\ti \x^\pm_{#1}}}
\newcommand{\mnt}[1]   {m_{\ti \x^0_{#1}}}

\def\tW              {\t_W}
\def\sth             {\sin\t}
\def\cth             {\cos\t}
\def\sthq            {\sin^2\t}
\def\cthq            {\cos^2\t}

\def\Emiss           {E\llap/}
\def\allsf           {$\st_1$, $\sb_1$, and $\stau_1$ }
\def\allpair         {$\stst$, $\sbsb$, and $\staustau$ }

\def\onehf               {\small{\frac{1}{2} }} 
\def\oneth               {\small{\frac{1}{3} }}    
\def\twoth               {\small{\frac{2}{3} }}

\def\rzw             {\sqrt{2}}

\def\ra              {\rightarrow}

\def\BR              {\rm BR}
\def\gev             {\:{\rm GeV}}
\def\pb              {${\rm pb}^{-1}$}

\def\leerz           {\hspace{1cm}\\}

\newcommand{\gsim}{\;\raisebox{-0.9ex}
           {$\textstyle\stackrel{\textstyle >}{\sim}$}\;}
\newcommand{\lsim}{\;\raisebox{-0.9ex}{$\textstyle\stackrel{\textstyle<}
           {\sim}$}\;}

\begin{titlepage}

\begin{flushright}
UWThPh-1996-18 \\
HEPHY-PUB 642/96\\
hep-ph/9603410\\
\vspace{0.2cm}
March, 1996
\end{flushright}

\bce

\vspace{1cm}
\begin{LARGE} {\bf
  Production of \\[2mm]
  Stop, Sbottom, and Stau  \\[2mm]
  at LEP2 \\[2mm]
}\end{LARGE}

\vspace{2cm}

\begin{large} 
A. Bartl$^{\small 1}$, 
H. Eberl$^{\small 2}$, \\[1mm]
S. Kraml$^{\small 2}$, 
W. Majerotto$^{\small 2}$, 
W. Porod$^{\small 1}$ \\[5mm]
\end{large}

{\em (1) Institut f\"ur theoretische Physik, Univ. Vienna, Austria} \\[1mm]
{\em (2) Institut f\"ur Hochenergiephysik (HEPHY), Vienna, Austria} 

\vfill

\begin{abstract}
We present a comprehensive study of 
pair production and decay of stops, sbottoms,
and staus in $e^+ e^-$ annihilation at LEP2. We give numerical predictions 
within the Minimal Supersymmetric Standard Model for cross sections and 
decay rates, and discuss the important signatures. In the case of stau 
production we also study the polarization of the $\tau$ in the decays
$\tilde{\tau}_1 \to \tau \tilde\chi^0_{1,2}$.
\end{abstract}

\ece 
\end{titlepage}

\baselineskip=21pt   

\section {Introduction}

Supersymmetry (SUSY) \cite{susy} requires the existence of two scalar
partners $\sq_{L}$ and $\sq_{R}$ (squarks) for every quark corresponding 
to its two helicity states.
For the same reason every lepton $\ell$ has as supersymmetric partners
the sleptons $\slep_{L}$ and $\slep_{R}$.
Whereas in the case of the first and second generation the 
$\sq_{L},\,\sq_{R}$ and $\slep_{L},\,\slep_{R}$ states are to a good 
approximation also the mass eigenstates, this is not expected for the 
third generation. 
Quite generally, $\sq_{L}$ and $\sq_{R}$ ($\slep_{L}$ and $\slep_{R}$) 
mix, the size of the mixing being proportional to the mass of the 
quark $q$ (lepton $\ell$). Thus it may be that one of the mass eigenstates of 
the stops, $\st_{1}$, is the  lightest squark. It could even be possible
that it is the lightest visible SUSY particle.
If the parameter $\tan\b$, which also enters the mixing, is large 
($\tan\b \gsim 10$), also the sbottom $\sb_{1}$ or the stau $\stau_{1}$
can be relatively light. 
Therefore, it is interesting to study stop, sbottom, and stau production
in the energy range of LEP2, that is $m_{Z} \leq \sqrt{s} \leq 192.5$ GeV.
The present experimental bound from LEP1 for the masses of charged
supersymmetric particles is $\ti m \gsim 45$~GeV \cite{lep1, grivaz}. 
Stronger limits for the stop mass, up to 55~GeV, are reported from LEP at
$\sqrt{s} = 130 - 140$~GeV \cite{lep15}.
The D$\emptyset$ experiment at FNAL obtained
additional mass bounds for the stop \cite{fnal, grivaz} excluding the 
mass range $40\gev\lsim\mst{1}\lsim 100$ GeV if the mass difference 
$(\mst{1}-\mnt{1}) \gsim 30$ GeV, where the neutralino $\nt_{1}$ is 
the lightest supersymmetric particle (LSP). 

Here we shall calculate the production rates for $\eeto\stst$,
$\eeto\sbsb$, and $\eeto\staustau$ in the whole accessible mass range 
of LEP2 as a function of the mixing angle. SUSY-QCD corrections as 
well as initial state radiation (ISR) will be included.
We also discuss in detail the decay patterns of these particles.
The phenomenology of the decays of $\st$, $\sb$, and $\stau$
is different from that of the other squarks and sleptons due to their
non-negligible Yukawa couplings.

The framework of our calculations is the Minimal Supersymmetric 
Standard Model (MSSM) \cite{susy}. 
The parameters which determine the phenomenology of stops, sbottoms, 
and staus are $M$, $M^{\prime}$, the soft-breaking SU(2) and U(1) 
gaugino masses, $\mu$, the higgsino mass parameter, 
$\tan\b = v_{2}/v_{1}$, where $v_{1}$ and $v_{2}$ are the vacuum 
expectation values of the two Higgs doublets, $m_{A}$, the mass of the 
pseudoscalar Higgs boson $A^{0}$, and $M_{\ti F}$ 
$(\ti F = \ti Q, \ti U, \ti D, \ti L, \ti E)$ and $A_{f}$ 
$(f = t,\,b,\,\tau)$, the soft-breaking parameters which enter the mass
matrices of the stops, sbottoms, and staus.
We shall use the GUT relations 
$M^\prime = {\small\frac{5}{3}}\,M\,\tan^2\tW \sim 0.5 M$ and 
$m_{\ti g} = \frac{\a_s}{\a_w}\,M\,\sin^2\tW \sim 0.3 M$, 
where $m_{\ti g}$ is the gluino mass. 

In the energy range of LEP2 the most important decay modes of 
$\st_{1}$, $\sb_{1}$, and $\stau_{1}$ are 
$\st_{1}\to c\,\nt_{1},\,b\,\chp_{1}$, $\sb_{1}\to b\,\nt_{1},\,b\,\nt_{2}$,
and $\stau_{1}\to \tau\,\nt_{1,2},\,\nu_{\tau}\chm_{1}$ 
(assuming $\mstau{1}< m_{\ti\nu,\slep}$). 
Here $\ch_{1}$ is the lighter of the two charginos present in the MSSM
and $\nt_{1}$ is the lightest of the four neutralinos 
(with $\mnt{1}<\mnt{2}<\mnt{3}<\mnt{4}$).
It is important to note that the masses and couplings of the charginos
and neutralinos only depend on $M$, $\mu$, and $\tan\b$.

In the next section we present the formulae for the sfermion mixing,
the production cross sections and the decays of 
$\st_{1}$, $\sb_{1}$, and $\stau_{1}$, taking into account the mixing 
and the Yukawa couplings. In section~3 we give numerical results for 
the cross sections of $\eeto\stst ,\: \sbsb ,\: \staustau$ for 
various masses as a function of the mixing angle,
a numerical analysis of the decays of $\st_{1}$, $\sb_{1}$, and 
$\stau_{1}$ into charginos and neutralinos, as well as a discussion of 
the main signatures. Section~4 contains a short summary and conclusions.

\section {Production cross section and decay formulae}


The mass matrix for stops, sbottoms and staus in the $(\sf_L,\:\sf_R)$ 
basis (with $\sf = \st,\,\sb,\,\stau$) has the following form 
\cite{ellis}:

\beq
  {\cal M}^2_{\sf} = \left(\baa{ll} m_{\sf_L}^{2} & a_f m_f \\ 
                                    a_f m_f  & m_{\sf_R}^{2} \eaa\right) 
\eeq
with 
\beq \begin{array}{ll}
  \mst{L}^{2} = M_{\ti Q}^2 + m_t^2 + m_Z^2\cos 2\b\,
               (\onehf - \twoth\sin^2\tW ) , \hspace{2mm} &
  \mst{R}^{2} = M_{\ti U}^2 + m_t^2 - \twoth m_Z^2\cos 2\b\,\sin^2\tW , \\
  \msb{L}^{2} = M_{\ti Q}^2 + m_b^2 - m_Z^2\cos 2\b\,
               (\onehf - \oneth\sin^2\tW ) , \hspace{2mm} &
  \msb{R}^{2} = M_{\ti D}^2 + m_b^2 + \oneth m_Z^2\cos 2\b\,\sin^2\tW ,\\
  \mstau{L}^{2} = M_{\ti L}^2 + m_\tau^2 - m_Z^2\cos 2\b\,
                     (\onehf - \sin^2\tW ) , \hspace{2mm} &
  \mstau{R}^{2} = M_{\ti E}^2 + m_\tau^2 + m_Z^2\cos 2\b\,\sin^2\tW , 
\end{array} \eeq
and
\beq
  a_{t}m_{t} = m_{t}(A_{t} - \mu\cot\b ), \hspace{4mm}
  a_{b}m_{b} = m_{b}(A_{b} - \mu\,\tan\b ), \hspace{4mm}
  a_{\tau}m_{\tau} = m_{\tau}(A_{\tau} - \mu\,\tan\b ). 
  \label{eq:offdiag}
\eeq
\noi The mass eigenstates $\sf_1$ and $\sf_2$ are related to 
$\sf_L$ and $\sf_R$ by:
\beq
  {\sf_1 \choose \sf_2} = 
    \left(\baa{ll} \cth_{\sf} & \sth_{\sf} \\ 
                  -\sth_{\sf} & \cth_{\sf} \eaa\right)\:
    {\sf_L \choose \sf_R} 
  \label{eq:mixing}
\eeq
with the eigenvalues 
\beq
  m_{\sf_{1,2}}^2 = \onehf\, (m_{\sf_L}^2 + m_{\sf_R}^2) \mp 
         \onehf \sqrt{(m_{\sf_L}^2 - m_{\sf_R}^2)^2 + 4\,a_f^2 m_f^2}.
  \label{eq:masses}
\eeq
The mixing angle $\theta_{\sf}$ is given by
\beq {\small 
  \cth_{\sf} = - a_f m_f\, \sqrt{   
      \frac{1}{(m_{\sf_L}^2-m_{\sf_1}^2)^2 + a_f^2 m_f^2}},
  \hspace{5mm} 
  \sth_{\sf} = \sqrt{ \frac{(m_{\sf_L}^2-m_{\sf_1}^2)^2}
                              {(m_{\sf_L}^2-m_{\sf_1}^2)^2 + a_f^2 m_f^2}}.
} \label{eq:mixangl} \eeq
Hence, $|\cth_{\sf}| > 1/\rzw$ if $\msf{L}<\msf{R}$ and 
$|\cth_{\sf}| < 1/\rzw$ if $\msf{R}<\msf{L}$. \\


The reaction $\eeto\sfsf$ proceeds via $\gamma$ and $Z^{0}$ exchange.
The tree level cross section at a center-of-mass energy of $\sqrt{s}$ 
is given by:
\beq
  \s^{tree} = \frac{\pi\a^2 N_C}{3s}\,\b^3 \left[\, Q_f^2 
    + \left(
      \frac{(v_e^2+a_e^2)\,v_{\sf_1}^2}{16\,s^4_W c^4_W} \,s^{2}
      - \frac{Q_f\,v_e\,v_{\sf_1}}{2\,s^2_W c^2_W} \,s(s-m_Z^2)
    \right)
    \frac{1}{(s-m_Z^2)^2 + \Gamma_Z^2 m_Z^2} \,\right]\,
  \label{eq:sigtree}
\eeq
where $s^2_W = 1 - c^2_W = \sin^2\tW$, $v_e = 2\sin^2\tW - \onehf$ 
and $a_e = -\onehf$. 
$N_C$ is a colour factor which is 3 for squarks and 1 for sleptons. 
The $Z^0$ coupling to $\sfsf$ is proportional 
$v_{\sf_1} = 2\,(I^3_f\cos^2\t_{\sf}-Q_f\sin^2\tW)$.
Here $I^3_f$ and $Q_{f}$ are the third component of the weak isospin
and the charge of the fermion $f$ ($Q_{e} = -1$). 
$\s^{tree}$ shows the typical $\beta^{3}$ suppression where
$\b = \left(1 - 4\,m_{\sf_1}^2/s \right) ^{1/2}$
is the velocity of the outgoing scalar particles.
The interference of the $\gamma$ and $Z^{0}$ exchange 
contributions leads to a characteristic minimum of the cross section 
at 
\beq
  \cos^{2}\theta_{\sf}{\big |}_{min} = 
    \frac{Q_{f}}{I^{3}_{f}}\sin^2\tW\,
    \left[ 1 + (1-\frac{s}{m_{Z}^{2}})\cos^{2}\tW
           \frac{L_{e}+R_{e}}{L_{e}^{2}+R_{e}^{2}} \,\right]
\eeq
where $L_{e}=\sin^2\tW - \onehf$ and $R_{e}=\sin^2\tW$. 
The angular distribution has the familiar $\sin^2\vartheta$ shape,
with $\vartheta$ the scattering angle:
\beq
\frac{{\rm d}\,\s^{tree}}{{\rm d} \cos\vartheta} = 
                    {\small \frac{3}{4}} \,\sin^2\vartheta\, \s^{tree}.
\eeq

\noi In the case of squarks QCD radiative corrections are important. 
The conventional QCD corrections were calculated in 
\cite{drees, beenakker} including the radiation of soft and hard 
gluons (in ${\cal O}(\a_{s})$).
The QCD corrections within the MSSM including virtual gluino and 
squark exchange were computed in \cite{susyqcd}
(The corrections due to the four-squark interaction is zero in the
renormalization scheme used). 
In our numerical 
calculations we have included both the gluonic corrections and those 
due to gluino and squark exchange. The corrections due to the 
exchange of supersymmetric particles are, however, small in the energy 
range of LEP2.
Moreover, we have taken into account initial state radiation (ISR) \cite{isr}.
 

The sfermion interaction with neutralinos and charginos is given by 
\cite{squarks}:
\beq
  {\cal L} = 
  g\,\bar{f}\, (a_{ik}^{f}\,P_{L} + b_{ik}^{f}\,P_{R})\,\nt_{k}\,\sf_{i} + 
  g\,\bar{f}^{\prime}\, 
               (l_{ij}^{f}\,P_{L} + k_{ij}^{f}\,P_{R})\,\chp_{j}\,\sf_{i} +
  {\rm h.c.}
  \label{eq:Lint}
\eeq
with
\beq
  {a_{1k}^{f} \choose a_{2k}^{f}} = 
    \left(\baa{ll} \cth_{\sf} & \sth_{\sf} \\ 
                  -\sth_{\sf} & \cth_{\sf} \eaa\right)\:
  {h_{Lk}^{f} \choose f_{Rk}^{f}},
  \hspace{4mm}
  {b_{1k}^{f} \choose b_{2k}^{f}} = 
    \left(\baa{ll} \cth_{\sf} & \sth_{\sf} \\ 
                  -\sth_{\sf} & \cth_{\sf} \eaa\right)\:
  {f_{Lk}^{f} \choose h_{Rk}^{f}},  
  \label{eq:copmix}
\eeq

\begin{small}

\noi
\beq \begin{array}{ll}
  h_{Lk}^{t} = Y_{t} \left( 
               N_{k3}\sin\b - N_{k4}\cos\b \right),  &
  f_{Lk}^{t} = -\frac{2\sqrt{2}}{3} \sin\tW N_{k1} - \sqrt{2}\,
        (\onehf - \twoth\sin^{2}\tW ) \frac{N_{k2}}{\cos\tW}, \\
  h_{Rk}^{t} = Y_{t} \left( 
               N_{k3}\sin\b - N_{k4}\cos\b \right),  &
  f_{Rk}^{t} = -\frac{2\sqrt{2}}{3} \sin\tW 
               (\tan\tW N_{k2} - N_{k1}), 
  \label{eq:stntcop}
\end{array} \eeq

\noi
\beq \begin{array}{ll}
  h_{Lk}^{b} = -Y_{b} \left( 
               N_{k3}\cos\b + N_{k4}\sin\b \right), &
  f_{Lk}^{b} = \frac{\sqrt{2}}{3}\sin\tW N_{k1} +
     \sqrt{2}\,(\onehf - \oneth\sin^{2}\tW ) \frac{N_{k2}}{\cos\tW}, 
     \hspace{5mm} \\
  h_{Rk}^{b} = -Y_{b} \left( 
               N_{k3}\cos\b + N_{k4}\sin\b   \right),  &
  f_{Rk}^{b} = \frac{\sqrt{2}}{3}\sin\tW 
                   (\tan\tW N_{k2} - N_{k1}),                             
  \label{eq:sbntcop}
\end{array} \eeq

\noi
\beq \begin{array}{ll}
  h_{Lk}^{\tau} = -Y_{\tau} \left(  
                  N_{k3}\cos\b + N_{k4}\sin\b \right),  &
  f_{Lk}^{\tau} = \sqrt{2}\sin\tW N_{k1} +
              \sqrt{2}\,(\onehf - \sin^{2}\tW ) \frac{N_{k2}}{\cos\tW}, 
              \hspace{7.5mm} \\ 
  h_{Rk}^{\tau} = -Y_{\tau} \left( 
                  N_{k3}\cos\b + N_{k4}\sin\b   \right),  &   
  f_{Rk}^{\tau} = \sqrt{2}\sin\tW (\tan\tW N_{k2} - N_{k1}),                                     
  \label{eq:slntcop}
\end{array} \eeq

\end{small}

\noi for the sfermion-fermion-neutralino interaction, and
\beq \begin{array}{ll}
 l_{1j}^{t} =       -V_{j1}\cth_{\st} +
              Y_{t}\,V_{j2}\sth_{\st}, \hspace{6mm} &
 k_{1j}^{t} = Y_{b}\,U_{j2}\cth_{\st}, \\
 l_{2j}^{t} =        V_{j1}\sth_{\st} + 
              Y_{t}\,V_{j2}\cth_{\st}, \hspace{6mm} &
 k_{2j}^{t} = -\,Y_{b}\,U_{j2}\sth_{\st}, 
  \label{eq:stchcop}
\end{array} \eeq
\beq \begin{array}{ll}
 l_{1j}^{b} =       -U_{j1}\cth_{\sb} +
              Y_{b}\,U_{j2}\sth_{\sb}, \hspace{4.8mm} &
 k_{1j}^{b} = Y_{t}\,V_{j2}\cth_{\sb},\\
 l_{2j}^{b} =        U_{j1}\sth_{\sb} + 
              Y_{b}\,U_{j2}\cth_{\sb}, \hspace{4.8mm} &
 k_{2j}^{b} = -\,Y_{t}\,V_{j2}\sth_{\sb}, 
  \label{eq:sbchcop}
\end{array} \eeq
\beq \begin{array}{ll}
 l_{1j}^{\tau} =          -U_{j1}\cth_{\stau} +
                 Y_{\tau}\,U_{j2}\sth_{\stau}, \hspace{3.6mm} &
 k_{1j}^{\tau} = 0,\hspace{21.5mm}\\
 l_{2j}^{\tau} =           U_{j1}\sth_{\sb} +
                 Y_{\tau}\,U_{j2}\cth_{\stau}, \hspace{3.6mm} &
 k_{2j}^{\tau} = 0, 
  \label{eq:slchcop}
\end{array} \eeq


\noi for the sfermion-fermion-chargino interaction. 
$N_{ij}$ is the $4\times 4$ unitary matrix diagonalizing the 
neutral gaugino-higgsino mass matrix in the basis $\ti\gamma$, 
$\ti Z^{0}$, $\ti H^{0}_{1}\cos\b - \ti H^{0}_{2}\sin\b$,
$\ti H^{0}_{1}\sin\b + \ti H^{0}_{2}\cos\b$ \cite{neutralinos}. 
$U_{ij}$ and $V_{ij}$ are the $2\times 2$ unitary matrices diagonalizing 
the charged gaugino-higgsino mass matrix \cite{charginos}. 
We choose a phase convention in which $N_{ij}$, $U_{ij}$, and $V_{ij}$ are real. 
$Y_{f}$ denotes the Yukawa coupling,
\beq
  Y_{t} = m_{t}/(\sqrt{2}\,m_{W}\sin\b), \hspace{6mm} 
  Y_{b} = m_{b}/(\sqrt{2}\,m_{W}\cos\b),\hspace{6mm}
  Y_{\tau} = m_{\tau}/(\sqrt{2}\,m_{W}\cos\b).
  \label{eq:yukcop}
\eeq

\noi The respective decay widths then are 
\beq
  \Gamma (\sf_{i}\to f\nt_{k}) =
  \frac{g^{2}\lambda^{\onehf}(\msf{i}^{2}, m_{f}^{2}, \mnt{k}^{2})}
       {16\pi\msf{i}^{3}} \,
  \left[ (a_{ik}^{2} + b_{ik}^{2}) (\msf{i}^{2} - m_{f}^{2} - \mnt{k}^{2}) -
         4 a_{ik}b_{ik}m_{f}\mnt{k} \right]
  \label{eq:ntwidth}
\eeq
and
\beq
  \Gamma (\sf_{i}\to f^{\prime}\ch_{j}) =
  \frac{g^{2}\lambda^{\onehf}(\msf{i}^{2}, m_{f^{\prime}}^{2}, \mch{j}^{2})}
       {16\pi\msf{i}^{3}}\,
  \left[ (l_{ij}^{2} + k_{ij}^{2}) 
         (\msf{i}^{2} - m_{f^{\prime}}^{2} - \mch{j}^{2}) -
         4\, l_{ij}k_{ij}m_{f^{\prime}}\mch{j} \right]
  \label{eq:chwidth}
\eeq
where $\lambda (x,y,z) = x^{2}+y^{2}+z^{2}-2xy-2xz-2yz$. \\

\section {Numerical results}  

\subsection {Stop $\st_{1}$}   

The total cross sections for the process $\eeto\stst$ at $\sqrt{s} = 175$ GeV
and $\sqrt{s} = 192.5$ GeV are shown in \fig{stprod} as a function of
$|\cth_{\st}|$ for several mass values of $\,\st_{1}$. 
At $\sqrt{s} = 192.5$ (175) GeV, for a stop mass of 80 GeV 
the cross section  reaches 0.35 (0.18) pb. 
Assuming an integrated luminosity of 300 (500) \pb 
at $\sqrt{s} = 192.5$ (175) GeV one can thus expect $\sim$ 60 to 105 
(50 to 90) $\stst$ events for $\mst{1} \simeq 80$ GeV.
Moreover, the cross section shows a clear dependence on the mixing angle 
for $\mst{1}\lsim 80$ GeV and $|\cth_{\st}| \gsim 0.6$. 
In this region cross section measurements should therefore allow
to determine the stop mixing angle once the mass of $\st_{1}$ is known.
The importance of radiative corrections is illustrated in
\fig{Xcorr}, where we show the conventional QCD 
corrections in ${\cal O}(\alpha_{s})$, the corrections due to 
gluino exchange as well as the ISR corrections   
at $\sqrt{s} = 192.5 $ GeV as a function of $\mst{1}$ for 
$\cth_{\st} = 0.7$. 
The gluonic corrections enhance the tree level cross section rising 
from 17 to 41\% for stop masses in the range of 45 to 85 GeV. 
The corrections due to the exchange of supersymmetric particles 
are $\lsim1.2\%$ for $m_{\ti g} = 200$ GeV and $\mst{2} = 250$ GeV
and depend on the stop mixing angle. 
Initial state radiation turns out to alter the tree level cross 
section from $\sim 1.2\%$ to $\sim -21\%$ for $\mst{1}$ = 45 to 85 GeV.


Assuming $\mst{1} < m_{\slep,\ti\nu}$ the main decay modes of 
$\st_{1}$ are $\st_{1}\to c\,\nt_{1}$ and $\st_{1}\to b\,\chp_{1}$. 
The latter decay has practically 100\% branching ratio 
if it is kinematically allowed.
As $\chp_1$ further decays into $\nt_1\, q \bar q^\prime$ or 
$\nt_1\, \ell \bar \nu_l$ the signature is 
two acoplanar $b$ jets accompanied by two leptons + large missing energy 
($\Emiss$), or single lepton + jets + $\Emiss$, or jets + $\Emiss$. 
Here the $b$ tagging technique can be used to extract the signal. 
Moreover, in this case the $\chp_1$ will most likely be observed first 
and the information from its decay properties can then be used to identify
the $\st_1$.
On the other hand, if $\mst{1} < m_{\chp_1} + m_b$ the f\/lavour changing decay
$\st_1 \to c\,\nt_1$ has practically 100\% branching ratio. 
The signature is then two acoplanar jets + $\Emiss$. 
Generally, for $\,\st_1\to c\,\nt_1$ the invisible energy is larger 
than in case of $\,\st_1\to b\,\chp_1$.
In \fig{stmodes} we show the domains of the $\,\st_1$ decay modes 
in the ($M,\:\mu$)~plane for $m_{\st_1} = 80$ GeV and $\tan\b = 2$ (there 
is a small stripe where the decay $\st_1\to c\,\nt_2$ is also possible).
If, however, $m_{\ch_1}+m_b > \mst{1} > m_{\slep (\ti\nu)} + m_b \,(+m_\ell)$ 
the decays $\st_1\to b\,\nu_\ell\bar{\slep}$ or $\st_1\to b\,\bar\ell\ti\nu_\ell$, 
proceeding via a virtual $\chp_1$ can compete with the decay into $c\,\nt_1$. 
In this case the signature is again $2b + 2\ell + \Emiss$. 


If the lifetime of $\st_{1}$ is longer than the typical  
hadronization time of ${\cal O}(10^{-23}\,s)$, i.~e. $\Gamma \lsim 0.2$ GeV, 
$\st_1$ hadronizes first into a colourless $(\st_1 \bar q)$ or $(\st_1 q q)$ 
bound state before decaying. 
This is generally expected in case of $\,\st_{1}\to c\,\nt_{1}$ and 
$\st_1\to b\,\nu_\ell\bar{\slep},\; b\,\bar\ell\ti\nu_\ell$ since 
these decays involve the electroweak coupling twice \cite{hikasa}.
However, also the width of the $\,\st_{1}$ decay into $b\,\chp_{1}$ can 
be smaller than the hadronization scale as illustrated in \fig{stwidth}.
Here we show the width of $\st_1\to b\,\chp_1$ as a function 
of $\cth_{\st}$ for $\mst{1} = 85$ GeV, $\mch{1} \simeq 60$ GeV, 
$\tan\b = 2$, and three scenarios with $M \ll |\mu |$, $|\mu | \ll M$,
and $M \sim |\mu |$ respectively. \\
If, as for the curves (a) and (b), $M$ is much smaller than $|\mu |$ 
the lighter chargino is gaugino-like.
In this case the $\st_{1}$-$b$-$\chm_{1}$ interaction is dominated by 
$V_{11}\,\cth_{\st}$ in \eq{stchcop} for $|\cth_{\st}| \gsim 0.3$. 
The deviation of the decay 
width from the $\cos^{2}\t_{\st}$ shape and the asymmetry in the sign 
of $\mu$ are due to constructive and destructive interferences with 
the term proportional to the top Yukawa coupling ($Y_{t}\,V_{12}\sth_{\st}$) 
which becomes especially important for $\st_{1}\sim\st_{R}$ 
and increases with decreasing $|\mu|$. \\
On the other hand, a higgsino-like $\chp_{1}$ strongly couples to $\st_R$, 
the dominant coupling being proportional to the top Yukawa coupling
($|\mu | \ll M$, curves (c) and (d)). Thus the decay width is large
if $|\cth_{\st}|\sim 0$ but goes to zero for $\st_{1}\sim\st_{L}$. \\
For $M \sim |\mu |$, which is shown in (e) and (f), a complicated 
interplay of gaugino and higgsino couplings gives rise to an 
intricate dependence on the mixing angle and a large asymmetry in the
sign of $\mu$.

The process of fragmentation of $\st_{1}$ was discussed in detail in 
ref.~\cite{beenakker}. Fast moving stops first radiate off gluons at small
angles. This process can be treated perturbatively. After that the 
(non--perturbative) hadronization phase follows by forming $(\st_{1}\bar q)$
and $(\st_{1} q q)$ hadrons. For $\beta \sim 1/2$ the energy loss of $\st_{1}$
due to gluon radiation and due to hadronization is of comparable size. Near
the threshold, the gluon emission suffers a $\b^{4}$ suppression.


Monte Carlo studies for $\st_{1}$ pair production have been performed within
the CERN--LEP2 Workshop 1995 \cite{lep2ws}. They have mainly 
concentrated on the decay $\st_{1}\to c\,\nt_{1}$ 
since in this case $\st_{1}$ would most probably be the first SUSY particle 
to be discovered. 
For the simulation of $\st_{1}$ fragmentation different approaches 
have been used. 
The conclusions have been a $5\sigma$ discovery reach of $\mst{1}\simeq$ 
75 to 90 GeV and a 95\% CL exclusion of $\mst{1}\simeq$ 84 to 92 GeV 
at $\sqrt{s} = 190$ GeV, for ${\cal L} = 300$ \pb 
depending on the stop mixing angle and the mass of the lightest 
neutralino. In case of $\st_{1}\to b\,\chp_{1}$ the experimental
reach for $\mst{1}$ is $\sim$ 85 GeV.

\subsection {Sbottom $\sb_{1}$}   

According to \eq{offdiag} a considerable $\sb_{L}$-$\sb_{R}$ mixing 
is possible if $\tan\b$ is large ($\tan\b > 10$). In this case the 
sbottom $\sb_{1}$ can be rather light. 
Hence, it is also interesting to discuss $\sb_{1}$ search at LEP2. 

The total (SUSY-QCD \& ISR corrected) cross sections of $\sb_{1}$ 
pair production at $\sqrt{s} = 175$ GeV and $\sqrt{s} = 192.5$ GeV 
are shown in \fig{sbprod} as a function of $|\cth_{\sb}|$ for several mass 
values of $\sb_{1}$.  
As can be seen, the dependence on the mixing angle is even more pronounced 
than in case of $\st_{1}$ production. 
The cross section of $\eeto\sbsb$ is smaller than the $\stst$ cross section
by a factor of 4 to $\sim 5/4$ for $|\cth_{\sb}|$ in the range of 0 to 1.
Again, radiative corrections are important. Standard QCD and ISR corrections 
are similar to the stop case, i. e. gluon corrections enhancing
the tree level cross section from 17 to 41\% for $\sb_{1}$ masses in the range 
of 45 to 85 GeV and initial state radiation lowering it by up to 21\%.
The corrections due to gluino exchange are $\lsim 1\%$
for $m_{\ti g} = 200$ GeV and $\msb{2} = 250$ GeV.


Assuming $\msb{1}<m_{\ti g}$ the main decay modes of $\sb_1$ are 
$\sb_1\to b\,\nt_1$ and 
$\sb_1\to b\,\nt_2$, the second decay being possible in the parameter 
region approximately given by $M < m_{\sb_1}-m_b$ or $|\mu| < m_{\sb_1}-m_b$. 
For the $b\,\nt_1$ channel the signature is two acoplanar $b$ jets + 
missing energy $\Emiss$. If the $\sb_1$ decays into $b\,\nt_2$ the 
$b$ jets are acompanied by additional jets and/or leptons from 
$\nt_2\to\nt_1\,q\bar q$ and/or $\nt_2\to\nt_1\,\ell\bar\ell$. 
$b$ tagging will help to enhance the signal. 
The domains of the $\sb_{1}$ decays in the $(M,\,\mu)$ plane
are shown in \fig{sbmodes} for $\msb{1} = 80$ GeV and $\tan\b = 40$. 

In \fig{BRsbct} we show the branching ratio for $\sb_1\to b\,\nt_1$ 
as a function of $\cth_{\sb}$ for $m_{\sb_1} = 80$ GeV, $\tan\b = 40$, 
$|\mu| = 500$ GeV, and $M = 55$, 65, and 75 GeV. 
As can be seen, the branching ratio is highly dependent on the mixing 
angle, if $M\ll|\mu|$ and both $\nt_{1}$ and $\nt_{2}$ are light. In 
this case $\sb_{1}$ mainly decays into $b\,\nt_{1}$ if 
$|\cth_{\sb}|\sim 0$, whereas for $\sb_{1}\sim\sb_{L}$ the decay into
$b\,\nt_{2}$ dominates. 
Evidently, the dependence on the mixing angle weakens with increasing
$\nt_{2}$ masses.
If, however, $M\sim|\mu|$ or $M\gg|\mu|$ the decay $\sb_{1}\to b\,\nt_{1}$ 
is the dominant mode.
The $\mu$ dependence of the branching ratio is shown in \fig{BRsbmu} 
for $m_{\sb_1} = 80$ GeV, $|\cth_{\sb}| = 0.72$, $\tan\b = 40$, and 
several values of $M$. 
As the parameter $A_{b}$ should not become unnaturally large 
in supergravity models \cite{sugra}, for large values of $\tan\beta$,
the parameter $\mu$ determines the sign of $\cth_{\sb}$
(see \eq{offdiag} and \eq{mixangl}).
For this reason, we have chosen $\cth_{\sb}$ and $\mu$ such that they
have the same sign in \fig{BRsbct}, \fig{BRsbmu}, and \fig{sbwidth}.


As in the $\st_1$ case the decay width of $\sb_1$ can be signif\/icantly
smaller than 0.2 GeV as illustrated in \fig{sbwidth}. 
Here we show the total width of $\sb_1$ as a function 
of $\mu$ for $m_{\sb_1} = 80$ GeV, $|\cth_{\sb}| = 0.72$, $\tan\b = 40$, and 
several values of $M$.
The decay width only exceeds the hadronization scale ($\Gamma\gsim 0.2$ GeV)
if $\nt_{1}$ is very light ($\lsim$ 30 GeV) or if the light neutralinos have 
strong higgsino components (small $|\mu|$). 
It clearly increases with decreasing $|\mu|$ as then   
$\nt_{1,2}$ become lighter and the higgsino couplings proportional
to $Y_{b}$ gain importance. So does the dependence on the mixing 
angle, which shows minima at $\cth_{\sb} = \pm 1$ and $\cth_{\sb} = 0$ 
where one decay channel ($b\,\nt_{1}$ or $b\,\nt_{2}$) dominates.


DELPHI has studied $\sb_{1}$ search when $\sb_{1}$ decays into 
$b\,\nt_{1}$ \cite{lep2ws}. 
Their conclusion was that in this case the discovery potential for 
$\sb_{1}$ is similar to the one of $\st_{1}\to c\,\nt_{1}$, i. e. 
$\msb{1} \lsim 75$ to 90 GeV depending on $\t_{\sb}$ and $\mnt{1}$.

\subsection {Stau $\stau_{1}$}   

As in the sbottom sector one expects large $\stau_{L}$-$\stau_{R}$ 
mixing for high values of $\tan\b$. This would lead to a light $\stau_{1}$
which could also lie in the energy range of LEP2.

The $\staustau$ production cross sections are plotted in \fig{slprod}
at $\sqrt{s} = 175$ GeV and $\sqrt{s} = 192.5$ GeV as a function of 
$|\cth_{\stau}|$ for several $\stau_{1}$ masses. 
As can be seen, the dependence on the mixing angle is much 
weaker for $\staustau$ than for $\stst$ and $\sbsb$ production.
At $\sqrt{s} = 192.5$ GeV the cross section is of ${\cal O}(0.1)$ pb 
for $\mstau{1} = 80$ GeV and goes up to 0.55 pb for a $\stau_{1}$ mass 
of 50 GeV. 
This corresponds to a production rate of $\sim 30$ (25) events at 
$\sqrt{s} = 192.5$ (175) GeV for an integrated luminosity of 
${\cal L} = 300$ (500) \pb ~and $\mstau{1} = 80$ GeV. 
Of course, here only ISR corretions have to be taken into account.


The simplest signature of $\staustau$ production is $2\tau + \Emiss$ 
with the $\tau$ leptons coming from $\stau_1\to \tau\,\nt_1$ decays.
If the decay modes into $\tau\,\nt_{2}$ and $\ti\nu_{\tau}\,\chm_{1}$ 
are also kinematically accessible the signature is 
$2\tau$ + jets and/or leptons + $\Emiss$, or single $\tau$ + 
jets and/or leptons + $\Emiss$, or jets and/or leptons + $\Emiss$ 
due to cascade decays of $\nt_{2}$ and $\chm_{1}$. The latter case occurs 
when both staus decay via $\stau_{1}\to\ti\nu_{\tau}\,\chm_{1}$ 
($\bar{\stau}_{1}\to\bar{\ti\nu}_{\tau}\,\chp_{1}$). 
The parameter domains of the various $\stau_{1}$ decays in the 
$(M,\,\mu)$ plane are shown in \fig{slmodes} for $\mstau{1} = 80$~GeV 
and $\tan\beta = 40$.
If $\ti e$, $\ti \mu$, and/or $\ti\nu$ are lighter than 
$\stau_{1}$ then also three--body decays into these particles plus 
neutrino(s), lepton(s), or quark pairs are possible.


In \fig{BRslct} we show the branching ratios of $\stau_1$ decays into 
$\tau\,\nt_1$, $\tau\,\nt_2$, and $\nu_{\tau}\,\chm_1$ as a 
function of $\cth_{\stau}$ for $m_{\stau_1} = 80$ GeV, $\tan\b = 40$, 
(a) $M \ll |\mu|$, and (b) $|\mu| \ll M$.  
As in section 3.2, we have chosen $\cth_{\stau}$ and $\mu$ such that they have 
the same sign in order to avoid unnaturally large values of $A_{\tau}$.
%
In case (a), i. e. $M \ll |\mu|$, if $\cth_{\stau} \lsim 0.5$ the decay 
$\stau_{1}\to\tau\,\nt_{1}$ has $\gsim 80\%$ branching ratio as
$\stau_{R}$ does not couple to a gaugino--like $\nt_{2}$ or $\chm_{1}$.
On the other hand, the decays into $\tau\,\nt_2$ and 
$\ti\nu_{\tau}\,\chm_{1}$ play an important r\^ole for 
$\stau_{1}\sim\stau_{L}$, i. e. $\cth_{\sb} \sim 1$. 
In contrast to that, for higgsino--like light neutralinos and charginos, 
i. e. $|\mu|\ll M$ as shown in (b), BR($\stau_1\to \tau\,\nt_1$) and 
BR($\stau_1\to \ti\nu_{\tau}\,\chm_{1}$) are of comparable size 
for $\stau_{1}\sim\stau_{R}$, whereas for $\stau_{1}\sim\stau_{L}$ 
the $\stau_{1}$-$\nu_{\tau}$-$\chm_{1}$ coupling vanishes and 
the decay into $\tau\,\nt_1$ has $\sim 100\%$ branching ratio.
The $\stau_{1}\to\tau\,\nt_{2}$ decay mode is negligible in this case.


For staus a new interesting aspect comes into play: 
Due to the supersymmetric versions of gauge and Yukawa interactions 
the sfermion-fermion-gaugino interaction conserves chirality while 
the sfermion-fermion-higgsino interaction f\/lips it. The polarization 
of the f\/inal state fermion thus depends on the sfermion (L-R) mixing 
as well as on the gaugino--higgsino mixing.
As, in contrast to other leptons, taus decay in the detector, one can 
determine their average polarization through the energy 
distribution of their decay products \cite{taupol}. 
Hence, one has an additional opportunity to obtain information on the 
$\stau_L$--$\stau_R$ and the gaugino--higgsino mixing by measuring 
the average polarization of tau leptons coming from 
$\stau_{1}\to\tau\nt_{k}$ decays \cite{nojiri}. \\
The average polarization of the $\tau$ leptons is given by:
\beq
  \bar{\cal P}(\tau) = 
  \frac{\sum_{k}\, {\rm BR}(\stau_{1}\to\tau_{R}\,\nt_{k}) - 
                   {\rm BR}(\stau_{1}\to\tau_{L}\,\nt_{k})}
       {\sum_{k}\, {\rm BR}(\stau_{1}\to\tau_{R}\,\nt_{k}) + 
                   {\rm BR}(\stau_{1}\to\tau_{L}\,\nt_{k})}
  \label{eq:taupol}
\eeq
$\bar{\cal P}(\tau)$ depends on the stau mixing angle $\theta_{\stau}$ 
and on the mixing matrix $N_{ij}$ of the neutral gaugino--higgsino 
sector. If the lighter neutralinos are gaugino--like only $N_{k1}$ and 
$N_{k2}$ give sizeable contributions. Thus, in this case 
$\bar{\cal P}(\tau)\simeq +1$ for $|\cth_{\stau}|\simeq 0$, whereas     
$\bar{\cal P}(\tau)\simeq -1$ for $|\cth_{\stau}|\simeq 1$. 
On the other hand, in case of higgsino--like $\nt_{1}$ and $\nt_{2}$, 
one finds 
$\bar{\cal P}(\tau)\simeq -1$ for $|\cth_{\stau}|\simeq 0$ and 
$\bar{\cal P}(\tau)\simeq +1$ for $|\cth_{\stau}|\simeq 1$. 
This is illustrated in Fig.~12 where we show the average polarization of 
tau leptons arising from $\stau_{1}\to\tau\,\nt_{k}$ decays 
as a function of $\cth_{\stau}$ for $m_{\stau_1} = 80$ GeV, $\tan\b = 40$, 
and three scenarios of (a) $M\ll |\mu|$, (b) $|\mu|\ll M$, and (c) $M\sim |\mu|$. 
In case of $M\sim |\mu|$ where the lighter neutralinos have both, 
gaugino and higgsino, properties $\bar{\cal P}(\tau) \sim \onehf$ and 
varies only little with the stau mixing angle. 


OPAL has studied the experimental aspects of a stau search at LEP2 
for $\mstau{R} \ll \mstau{L}$ \cite{lep2ws}. At $\sqrt{s} = 190$ GeV and 
for ${\cal L} =$ 300 \pb ~they obtained a $5\sigma$ detectability for 
stau pairs of $\mstau{R} \simeq$ 70 to 83 GeV for neutralino masses 
in the range of 20 to 72 GeV.
However, they have neglected the interesting possibility of 
$\stau_{L}$-$\stau_{R}$ mixing. 
 
\section {Conclusions}

We have discussed the phenomenology  of stop, sbottom, and stau pair 
production and decays at LEP2 paying particular attention to the 
sfermion L-R mixing. 
Analytical formulae have been given for the sfermion mixing, 
for the cross sections of $\st_{1}$, $\sb_{1}$, and $\stau_{1}$ 
pair production in \ee~annihilation, and for the widths of the decays of 
these particles into neutralinos and charginos. 
We have presented numerical predictions for these production and decay
processes and analyzed their dependence on the SUSY parameters.
It has turned out that due to L-R mixing and non-negligible Yukawa
couplings the phenomenology of stops, sbottoms, and staus 
can be significantly different from that of first and second 
generation sfermions.
Moreover, we have shown that in case of $\st_{1}$ and $\sb_{1}$ 
hadronization effects can be important.
For $\stau_{1}$ we have also discussed the dependence of the average
polarization of taus arising from $\stau_{1}\to \tau\,\nt_{k}$ decays 
on $\stau_{L}$-$\stau_{R}$ and gaugino-higgsino mixing.

\section* {Acknowledgements}

We thank our colleagues from the ``New Particles'' working group of 
the 1995 CERN--LEP2 workshop for useful discussions.
This work was supported by the 
''Fonds zur F\"orderung der wissenschaftlichen
Forschung'' of Austria, project no. P10843-PHY.


\newpage
\section* {Figure captions}

\vspace{5mm}
\refstepcounter{figure} 
\noindent{\bf Figure~\arabic{figure}:}~Total 
cross section for $\eeto\stst$ in pb 
at $\sqrt{s} = 175$ GeV (dashed lines) and $\sqrt{s} = 192.5$ GeV 
(solid lines) as a function of the stop mixing angle for $\st_{1}$ masses of 
50, 60, 70, 80, and 90 GeV. 
\label{fig:stprod}  
  
\vspace{5mm}
\refstepcounter{figure}    
\noindent{\bf Figure~\arabic{figure}:}~Gluon, gluino-stop, initial state,
and total radiative corrections relative to the tree level cross 
secion of $\eeto\stst$ at $\sqrt{s}=192.5$ GeV as a function of $\mst{1}$ 
for $\cth_{\st}=0.7$. 
\label{fig:Xcorr}

\vspace{5mm}
\refstepcounter{figure}    
\noindent{\bf Figure~\arabic{figure}:}~Parameter 
domains in the $(M,\:\mu)$ plane 
for the various $\st_1$ decay modes, for $m_{\st_1}=80\gev$ and 
$\tan\b=2$. The grey area is excluded by LEP1.
\label{fig:stmodes}

\vspace{5mm}
\refstepcounter{figure}    
\noindent{\bf Figure~\arabic{figure}:}~Decay width of $\st_1\to b\,\chp_1$
for $\mst{1}=85\gev$, $m_{\chm_1} \simeq 60\gev$, $\tan\b = 2$, 
(a) $M = 51.2$ GeV, $\mu = -500$ GeV, 
(b) $M = 72.0$ GeV, $\mu = +500$ GeV, 
(c) $M = 400$ GeV,  $\mu = -49.4$ GeV, 
(d) $M = 400$ GeV,  $\mu = +75.5$ GeV, 
(e) $M = 50$ GeV,   $\mu = -47.9$ GeV, and 
(f) $M = 135$ GeV,  $\mu = +132$ GeV. 
\label{fig:stwidth}

\vspace{5mm}
\refstepcounter{figure}    
\noindent{\bf Figure~\arabic{figure}:}~Total cross section 
for $\eeto\sbsb$ in pb 
at $\sqrt{s} = 175$ GeV (dashed lines) and $\sqrt{s} = 192.5$ GeV 
(solid lines) as a function of the sbottom mixing angle for $\sb_{1}$ 
masses of 50, 60, 70, 80, and 90 GeV.
\label{fig:sbprod}

\vspace{5mm}
\refstepcounter{figure}    
\noindent{\bf Figure~\arabic{figure}:}~Parameter 
domains in the $(M,\:\mu)$ plane 
for the $\sb_1$ decay modes for $m_{\sb_1}=80\gev$ and 
$\tan\b=40$, (a) $\sb_{1}\to b\,\nt_{1}$ and 
(b) $\sb_{1}\to b\,\nt_{1,2}$. 
The grey area is excluded by LEP1.
\label{fig:sbmodes}

\vspace{5mm}
\refstepcounter{figure}    
\noindent{\bf Figure~\arabic{figure}:}~Branching ratio of 
$\sb_{1}\to b\,\nt_{1}$ 
in percent as a function of the sbottom mixing angle for $\msb{1} = 80$ GeV, 
$\tan\b = 40$, $|\mu| = 500$ GeV, and $M = 55$, 65, and 75 GeV.
\label{fig:BRsbct}

\vspace{5mm}
\refstepcounter{figure}    
\noindent{\bf Figure~\arabic{figure}:}~Branching ratio of 
$\sb_{1}\to b\,\nt_{1}$ 
in percent as a function of $\mu$ for $\msb{1} = 80$ GeV, $\cth_{\sb} = 0.72$,
$\tan\b = 40$, and $M = 55$, 65, and 75 GeV.
\label{fig:BRsbmu}

\vspace{5mm}
\refstepcounter{figure}    
\noindent{\bf Figure~\arabic{figure}:}~Total decay width of $\sb_{1}$ in GeV
as a function of $\mu$ for $\msb{1} = 80$ GeV, $\cth_{\sb} = 0.72$,
$\tan\b = 40$, and $M = 55$, 65, 75, and 150 GeV.
\label{fig:sbwidth}


\vspace{5mm}
\refstepcounter{figure}    
\noindent{\bf Figure~\arabic{figure}:}~Total cross section for 
$\eeto\staustau$ in pb 
at $\sqrt{s} = 175$ GeV (dashed lines) and $\sqrt{s} = 192.5$ GeV 
(solid lines) as a function of the stau mixing angle for $\stau_{1}$ masses of 
50, 60, 70, 80, and 90 GeV.
\label{fig:slprod}

\vspace{5mm}
\refstepcounter{figure}    
\noindent{\bf Figure~\arabic{figure}:}~Parameter domains in the 
$(M,\:\mu)$ plane 
for the various $\stau_1$ decay modes, for $m_{\stau_1}=80\gev$ and 
$\tan\b=40$ with (a) $\stau_{1}\to\tau\,\nt_{1}$, 
(b) $\stau_{1}\to\tau\,\nt_{1},\:\nu_{\tau}\chm_{1}$, and 
(c) $\stau_{1}\to\tau\,\nt_{1},\:\tau\,\nt_{2},\:\nu_{\tau}\chm_{1}$
The grey area is excluded by LEP1.
\label{fig:slmodes}

\vspace{5mm}
\refstepcounter{figure}    
\noindent{\bf Figure~\arabic{figure}:}~Branching ratio of 
$\stau_{1}\to \tau\,\nt_{1}$
(solid lines), $\stau_{1}\to \tau\,\nt_{2}$ (dashed lines), and 
$\stau_{1}\to \nu_{\tau}\,\chm_{1}$ (dashdotted lines) 
in percent as a function of the stau mixing angle for $\mstau{1} = 80$ GeV, 
$\tan\b = 40$, (a) $M = 65$ GeV, $|\mu | = 500$ GeV, 
and (b) $M = 300$ GeV, $|\mu | = 60$ GeV.
\label{fig:BRslct}

\vspace{5mm}
\refstepcounter{figure}    
\noindent{\bf Figure~\arabic{figure}:}~Average polarization of $\tau$ leptons 
arising from $\stau_{1}\to\tau\,\nt_{k}$ decays as a function of the
stau mixing angle for $\mstau{1} = 80$ GeV, $\tan\b = 40$, 
(a) $M = 65$ GeV,  $|\mu | = 500$ GeV (solid line), 
(b) $M = 300$ GeV, $|\mu | =  60$ GeV (dashdotted line), and
(c) $M = 100$ GeV, $|\mu | = 100$ GeV (dashed line).
\label{fig:slpol}

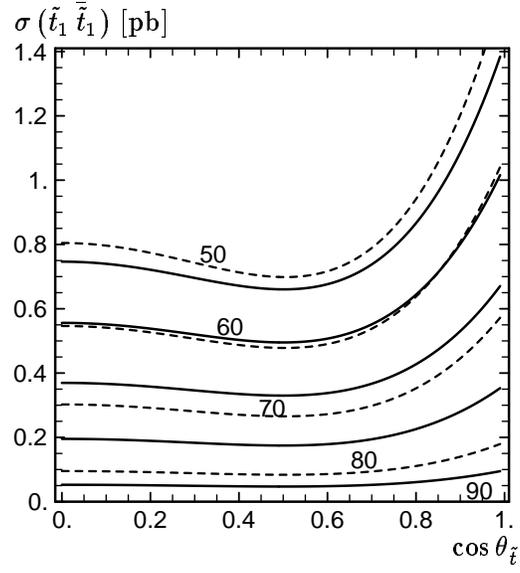
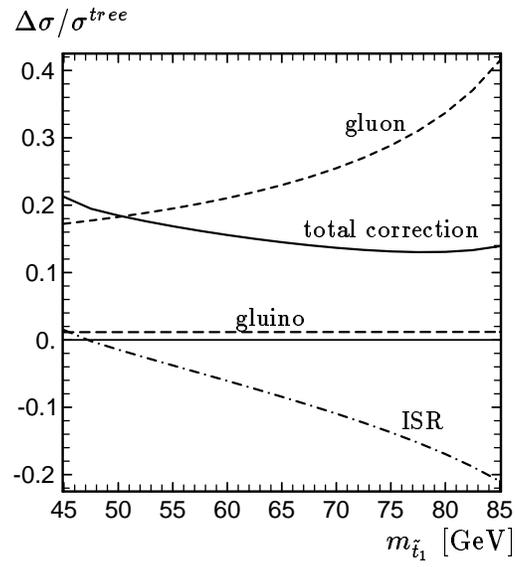
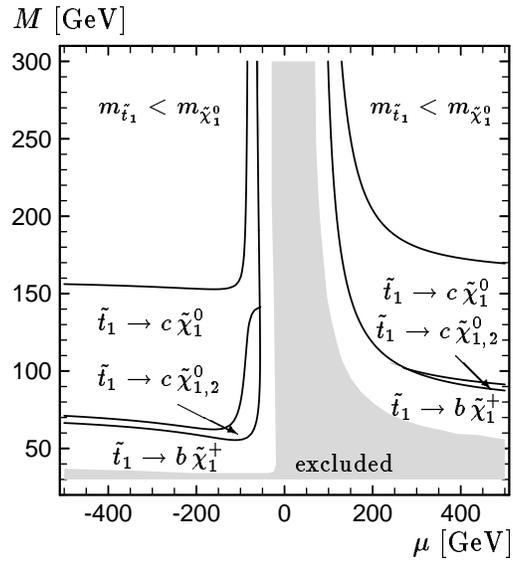
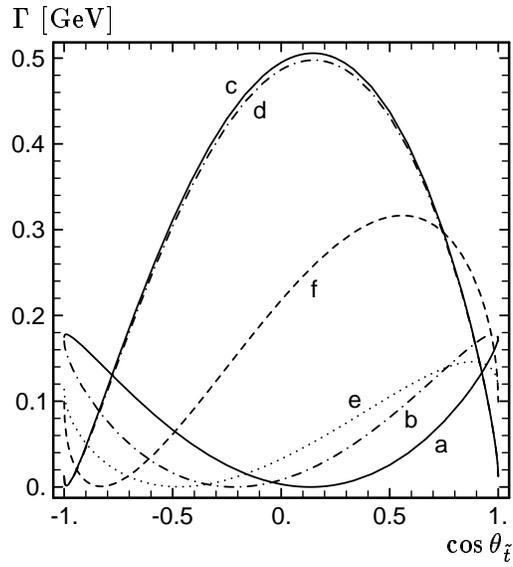
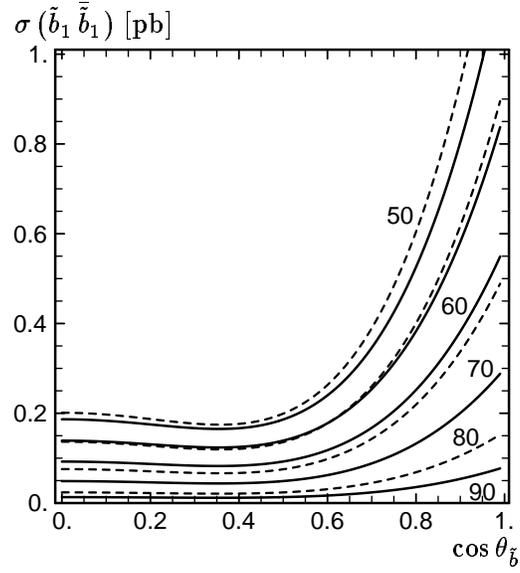
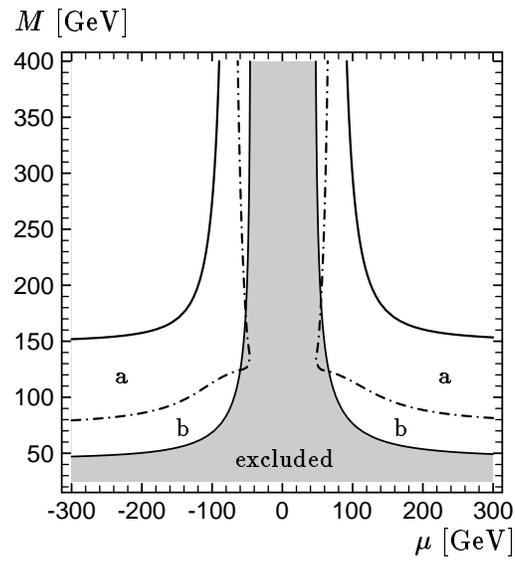
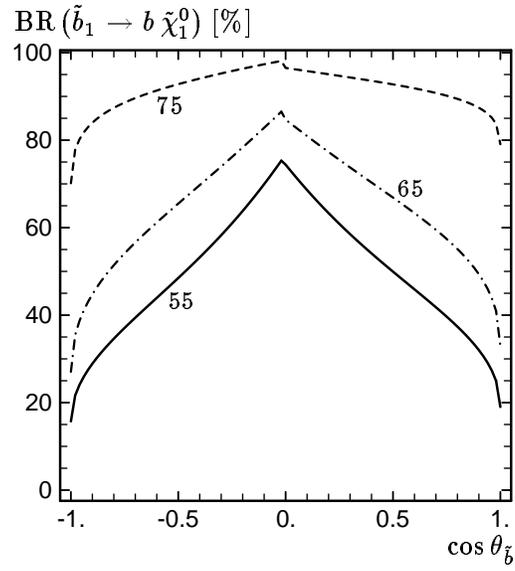
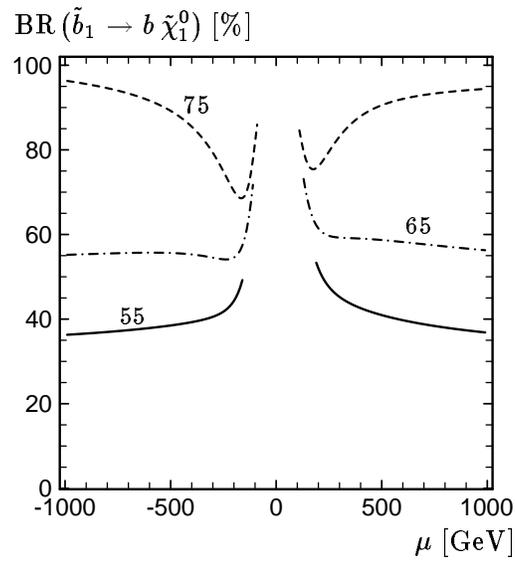
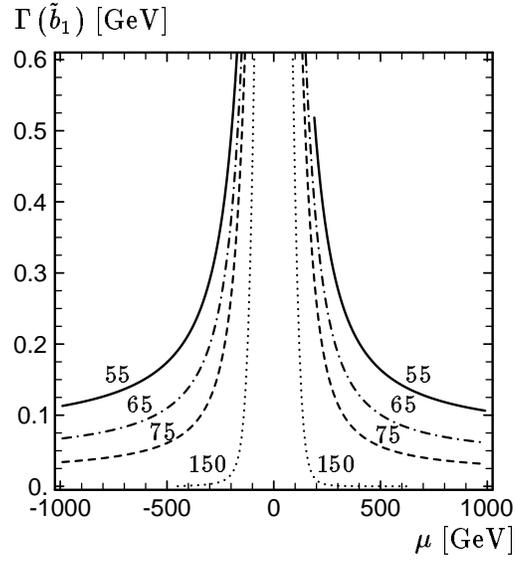
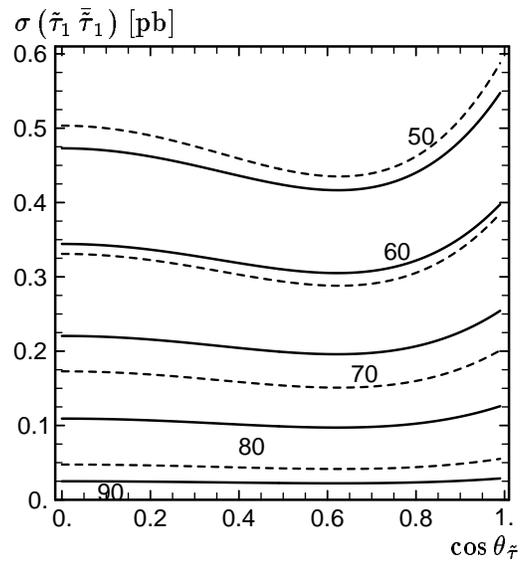
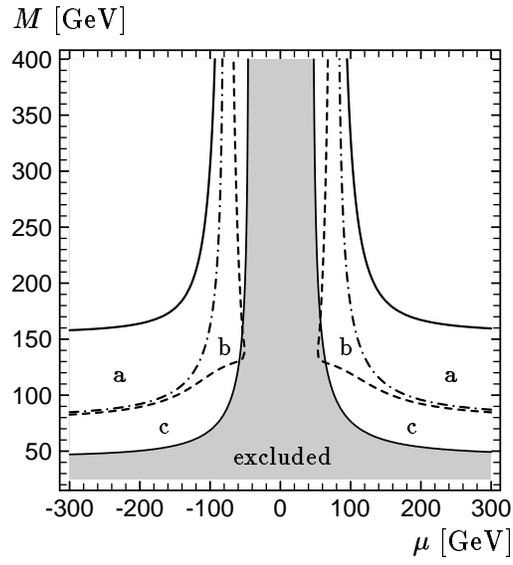
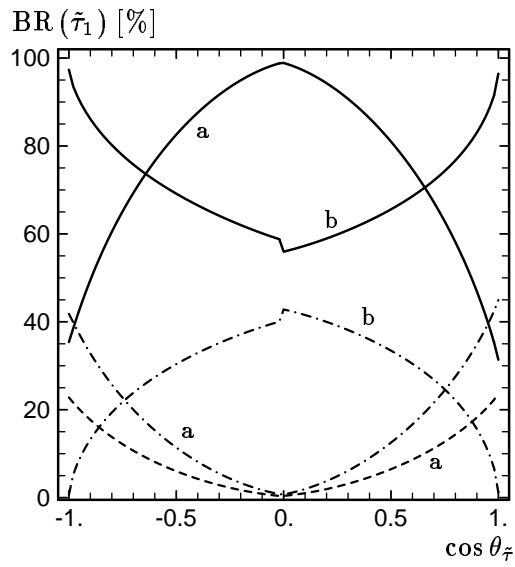
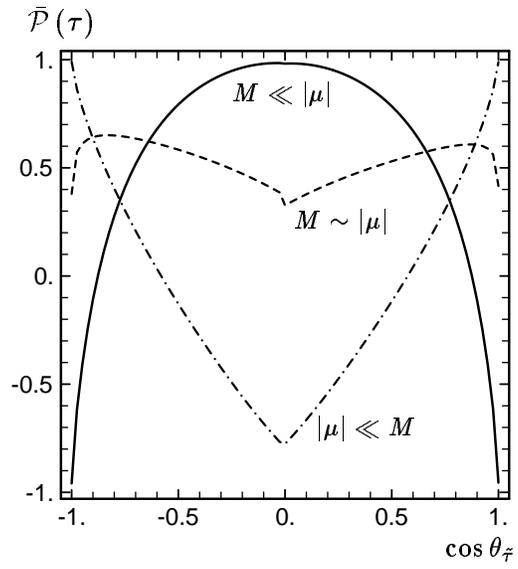
\end{document}